\documentclass[12pt]{article}
\usepackage{graphicx}

\def\deg{$^{\rm o}\;$}
\begin{document}

\title{Atmospheric hypotheses' of Earth's global warming}

\author{Vladimir Shaidurov\footnote{Institute of Computational Modeling of Siberian Branch of Russian
Academy of Sciences, shidurov$@$icm.krasn.ru.  Visited Leicester in
April 2005.}}

\date{}

\maketitle

\begin{abstract}
Two hypotheses are presented, outlining a new cause for global
warming. We propose that the crucial factor in global warming is
the amount and position of water vapour through the atmosphere.
The purpose of this report is to open the debate and to encourage
discussion among scientists. \end{abstract}

\medskip

 When analyzing the
mean-year trend of the Earth's surface temperature for the past 140
years (see for Fig.~1(a)) one can discern two sections of monotone
linear increase of temperature during two last industrial
centuries.\footnote{The best estimate for this trend for the last
100 years is that the global average surface temperature has
increased by $0.6 \pm 0.2$ \deg C.} The first one begins somewhere
in the period 1906-1909. The previous segment demonstrates a weak
decrease in the temperature trend, not increase (see for Figs.~
1(b), 4), though this includes the beginning of industrial period,
with a subsequent increase in production of greenhouse gases. If we
suppose that it is not only greenhouse gases that launched global
warming, then what could be an additional cause of the comparatively
sudden change of mean-year trend of the Earth's surface temperature?
We should look for a phenomenon of cosmic scale during this time
which could have given rise to beginning of global warming with a
significant probability. On the 30th June 1908 Tungus meteorite
crossed almost all of the atmosphere and exploded. Instrumental
measures with numerical modelling reconstruct an explosion of the
power of approximately by 15 Mt TNT at an altitude approximately of
10 km. Such an explosion could cause considerable stirring of the
high layers of atmosphere and change its structure.

\medskip

{\bf Hypothesis 1.}\ The Tungus meteorite considerably changed the
thermo-protective properties of the Earth's atmosphere and turned
out to be one of the agencies which launched global warming.

\medskip

    From our point of view, water vapor plays a more important role
     in Earth's thermal regime than other gases. The point is that
 additionally to the usual accumulation of heat because of increasing temperature,
 water vapour (as molecules of ${\rm H}_2{\rm O}$) possesses three more significant properties in the
 real range of pressure and temperature in the lower layers of atmosphere: liquid-gas transition,
 solid-liquid transition, and dissociation into ions (${\rm H}_2{\rm O} \rightarrow {\rm H}^+ + {\rm OH}^-$). At constant temperature
 each of these transitions consumes heat, and the opposite transitions give the same heat that
 is more often greater than energy due to moderate changes of temperature. Moreover, the
 considerable heat capacity of water in comparison with the atmospheric gases makes the
 World Ocean a very important accumulator of heat and source of vapour. Together, the complex of atmosphere,
 land surface, and ocean, assimilates 25\% of the solar radiation incident on the Earth.
 After all stages of reradiation 97\% of this energy goes out of Earth. Of the components of this complex,
 changes in amount of water vapor form the primary influence on assimilation of solar radiation.

For a rough estimate for the influence of increase or decrease of
primary assimilation of solar energy, let as look at the minimal
temperature reached at the Earth's surface: -89\deg in the
Antarctic. Thermal processes inside the Earth do not lead to
significant temperature differences at the surface in the absence
of solar heating. Other parts of the Earth's surface are at
greater distance from center and liquid core owing to the Earth's
shape. Having an average temperature of land surface and ocean of
about +15$^{\rm o}$, we see that the increase of assimilation of
solar energy by 1\% would entail an increase of temperature by
more than 4\deg in stationary regime.

    But the real dynamic regime intensifies this effect.
    The point is that the average increase of land and ocean temperature produces higher
    average absolute humidity. In its turn, this raises the assimilation ability of the
    atmosphere even at constant content of carbon dioxide. But increasing the average ocean
    temperature is responsible for lower water solubility of carbon dioxide, which then arrives
    in the atmosphere. Moreover, increase in land temperature is responsible for growth of bogs,
    at least in Northern Russia, due to the removal of permafrost deep down. The rise in area and
    activity of bogs leads to more active production of methane. Thus, a self-stimulated process
    was launched for the increase of average temperature of the Earth's surface. Therefore the rise
    of greenhouse gas concentration is more a consequence of warming but not a main reason.

\medskip

    {\bf Hypothesis 2:}\ The above mentioned variant of self-stimulated process
    (with a permanent rise of average absolute humidity, and resulting
    concentration of carbon dioxide and other greenhouse gases) was launched in 1908
    after the atmosphere reconstruction due to the Tungus meteorite.

\medskip

    Following a further examination of the curve in Figure 1(a), one can see oscillatory
    behavior in the period 1945 - 1976. What was happened in this time? The 16th July 1945
    saw the first explosion of a nuclear bomb, and this heralded the period of nuclear bomb tests.
    From the standpoint of atmosphere hypothesis the nuclear tests in the atmosphere are opposite to
    consequences of the Tungus meteorite.  When a nuclear charge explodes at the Earth's surface
    or in the atmosphere, the shock wave vents water vapor from the troposphere to the stratosphere
    through tropopause. For some period ($\approx$ 3 years) water vapor in the stratosphere and aerosol,
    and dust in the troposphere and stratosphere suffice for the defense of the Earth from solar
    radiation. But then all gradually settled, and global warming continued. All nuclear explosions
    above the ground and the sea together gave rise to tendency for decreasing the global temperature
    of Earth's surface. The last nuclear test in open atmosphere was on the 16th October 1980. All
    subsequent nuclear tests were under the ground or the sea, which does not generates the shock
    wave reaching the stratosphere. But the tendency of global warming recommenced earlier,
    approximately in 1977. Discordance between beginning of second period of warming and the
    finish of nuclear tests in the open atmosphere may be attributed to superposition with
    cyclic variations of temperature, which were apparent before 1908 too.

    Now let us raise a question about some regulation of the protective
    properties of the atmosphere. In the first instance we have the previous
    consideration of nuclear explosions in the open atmosphere. But this gave
    a comparatively short-term effect because of the gradual subsidence of ice
    in pearl clouds through the tropopause, due to their greater density when
    compared with air. However, clouds reflect the significant part of solar
    radiation. But in the mesosphere there is one more type of clouds, the so
    called silver clouds. They persist much longer. The distinction is the following.
     Almost the whole of the stratosphere and mesosphere consists of molecular oxygen
     O$_2$ and molecular nitrogen N$_2$. Also ozone O$_3$ is formed in comparatively small
     quantities with the help of solar radiation. The first distinction is the temperature
     gradient: temperature grows with altitude in the stratosphere, approximately from -55\deg to
     0\deg
     and diminishes in the mesosphere, from about 0\deg to -95\deg (see Fig. 2). The second
     distinction consists in the different pressure and density, which are several
     times less in the mesosphere than in the stratosphere. Therefore, water vapour in the
     troposphere (such as is formed during atmospheric nuclear tests), comes to a temperature
     below freezing point almost everywhere except its upper border. Thus it forms crystals
     having greater density than the ambient gas. In rapidly moving flows the crystals migrate
     down through the tropopause into the troposphere. High speeds and agitation do not end this
     process quickly; it may continue for months depending upon the tropospheric humidity in the
     test region.

But water vapour in the mesosphere is another matter. At a
pressure hundreds of times less than at atmospheric pressure at
sea level, the freezing point of water vapour shifts to a vastly
negative temperature without the intermediate liquid state (see
Fig. 3). Therefore there exists a sizable layer spanning the
higher part of the stratosphere and lower part of the mesosphere
where water is in the gas state. The mesospheric composition is
slightly distinct from the stratospheric one at significant, with
less density of gases. The gaseous state of water vapor has lower
density than the ambient gases (atomic masses of H$_2$O, O$_2$,
and N$_2$ equal 18, 32, and 28, respectively). Therefore it has
some tendency to move up in rapidly moving flows with some
stirring against the background of diffusion. When it migrates,
gas climbs to a temperature below freezing point, crystallizes and
migrates down. There it evaporates missing the liquid state, and
the process repeats. Thus, mesopause with a strongly negative
temperature of around -95\deg prevents water vapor leaving beyond
the upper bound of the mesosphere.

From these discussions the following idea results about some
deceleration and possible ceasing of global warming. With this
purpose it is possible to start reconstruction of the protective
layer in middle part of mesosphere, which consists of water vapor.
For this it is enough to transport and combust molecular hydrogen,
H$_2$ in the appropriate part of the mesosphere. The ambient
quantity of molecular oxygen and ozone is enough to generate water
vapor, which will be 9 times greater by weight than the
transported hydrogen. Along with this, the density of molecular
hydrogen is small enough in comparison with that of the ambient
gases over the region of transport that it induces lift. Therefore
hydrogen may serve as a means for transport itself.

Modern technical tools seem to be sufficient to realize such a
transport in small parts and to observe the consequences. In
addition, modern mathematical and computational modeling are at
level when more detailed quantitative estimates are possible both
for the immediate effects of such intervention as well as its
influence on global climate.

Note the obvious environmental safety of this suggestion since the
combustion of hydrogen in higher level of atmosphere can generate
water vapor only. Combustion of small amounts gives the
possibility of observing any change before it becomes too
dramatic.

{\bf Acknowledgements:} In conclusion I express my gratitude to
Professor J. Levesley and Professor A. Gorban for the invitation
to visit the University of Leicester and for fruitful work and
discussions.

\begin{figure}
\begin{centering}
\includegraphics[width=11cm,height=13cm]{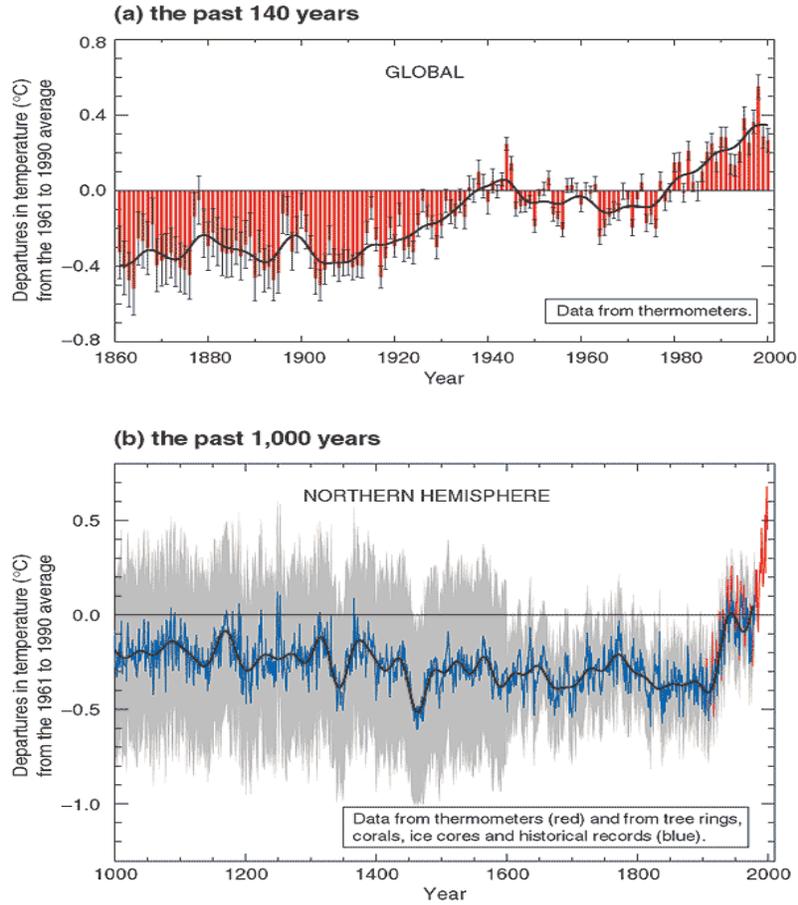}
\caption{\scriptsize (a) The Earth's surface temperature is shown
year by year (red bars) and approximately decade by decade (black
line, a filtered annual curve suppressing fluctuations below near
decadal time-scales). There are uncertainties in the annual data
(thin black whisker bars represent the 95\% confidence range) due to
data gaps, random instrumental errors and uncertainties,
uncertainties in bias corrections in the ocean surface temperature
data and also in adjustments for urbanization over the land.  (b)
Additionally, the year by year (blue curve) and 50 year average
(black curve) variations of the average surface temperature of the
Northern Hemisphere for the past 1000 years have been reconstructed
from ``proxy" data calibrated against thermometer data (see list of
the main proxy data in the diagram). The 95\% confidence range in
the annual data is represented by the grey region. These
uncertainties increase in more distant times and are always much
larger than in the instrumental record due to the use of relatively
sparse proxy data. Nevertheless the rate and duration of warming of
the 20th century has been much greater than in any of the previous
nine centuries. Similarly, it is likely that the 1990s have been the
warmest decade and 1998 the warmest year of the millennium. (The
figure is reprinted from  [1].)}
\end{centering}
\end{figure}

\begin{figure}
\begin{centering}
\includegraphics[width=11cm,height=16cm]{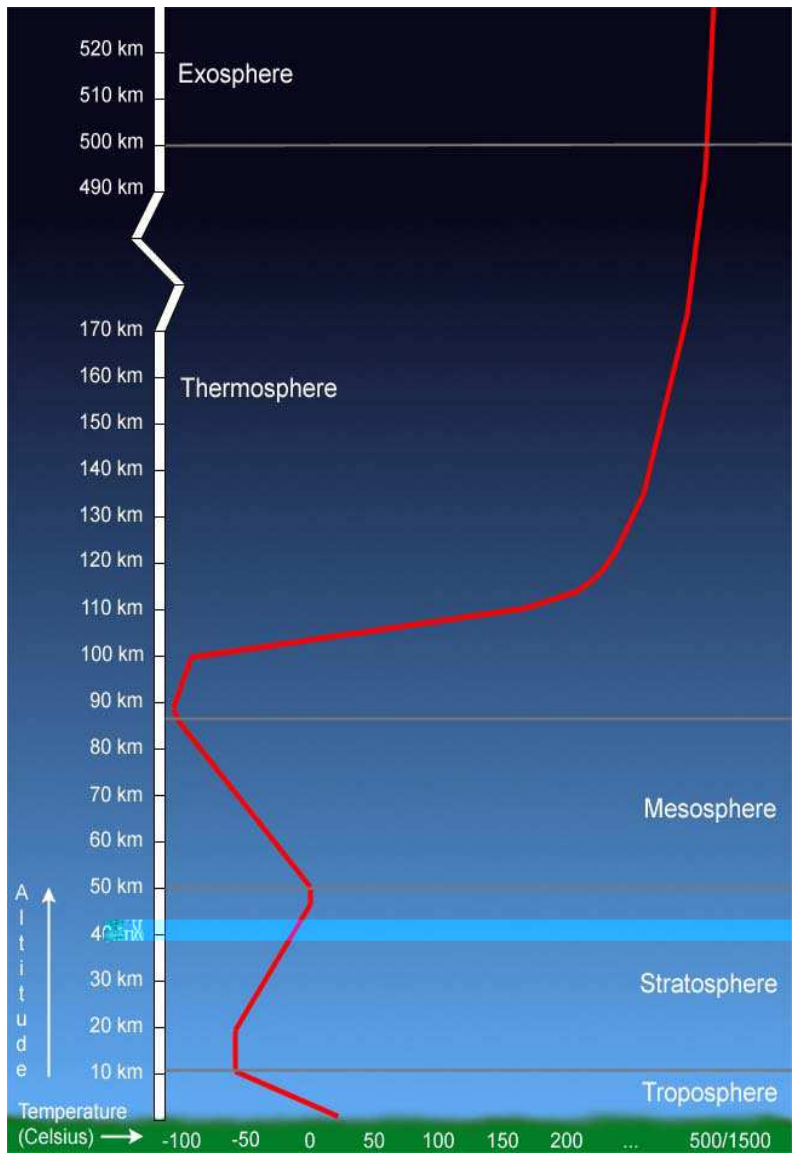}
\caption{The average temperature profile through Earth's atmosphere.
(The figure is reprinted from [2].) }
\end{centering}
\end{figure}

\begin{figure}
\begin{centering}
\includegraphics[width=12cm,height=12cm]{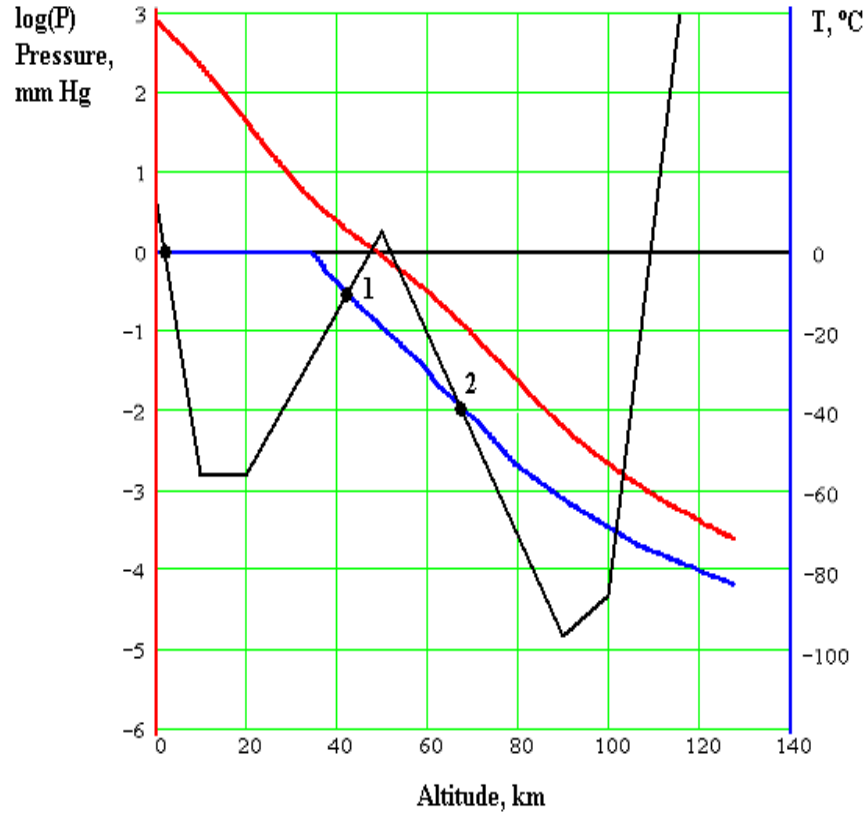}
\caption{The pressure, the freezing point of water vapor, and the
temperature of average atmosphere depending on altitude. (The data
are taken from [2], [3], [4].)  The abscissa is the altitude in km.
The pressure is plotted at left ordinate as logarithm of mercury
column in mm. The temperature is plotted at right ordinate in
Celsius degree. The red line gives an estimate of the pressure at
corresponding altitude. The blue line gives the freezing point of
water vapor depending on pressure (at corresponding point). The
black line demonstrates the temperature of average atmosphere at
corresponding altitude and copies the analogous graph from Figure 2
in some different form. The node of intersection 1 indicates the
point of formation for pearl clouds; the node 2 does for silver
ones. Between them the water vapor is in gas state. Near above and
near below it is in solid state.  }
\end{centering}
\end{figure}

\begin{figure}
\begin{centering}
\includegraphics[width=15cm,height=19cm]{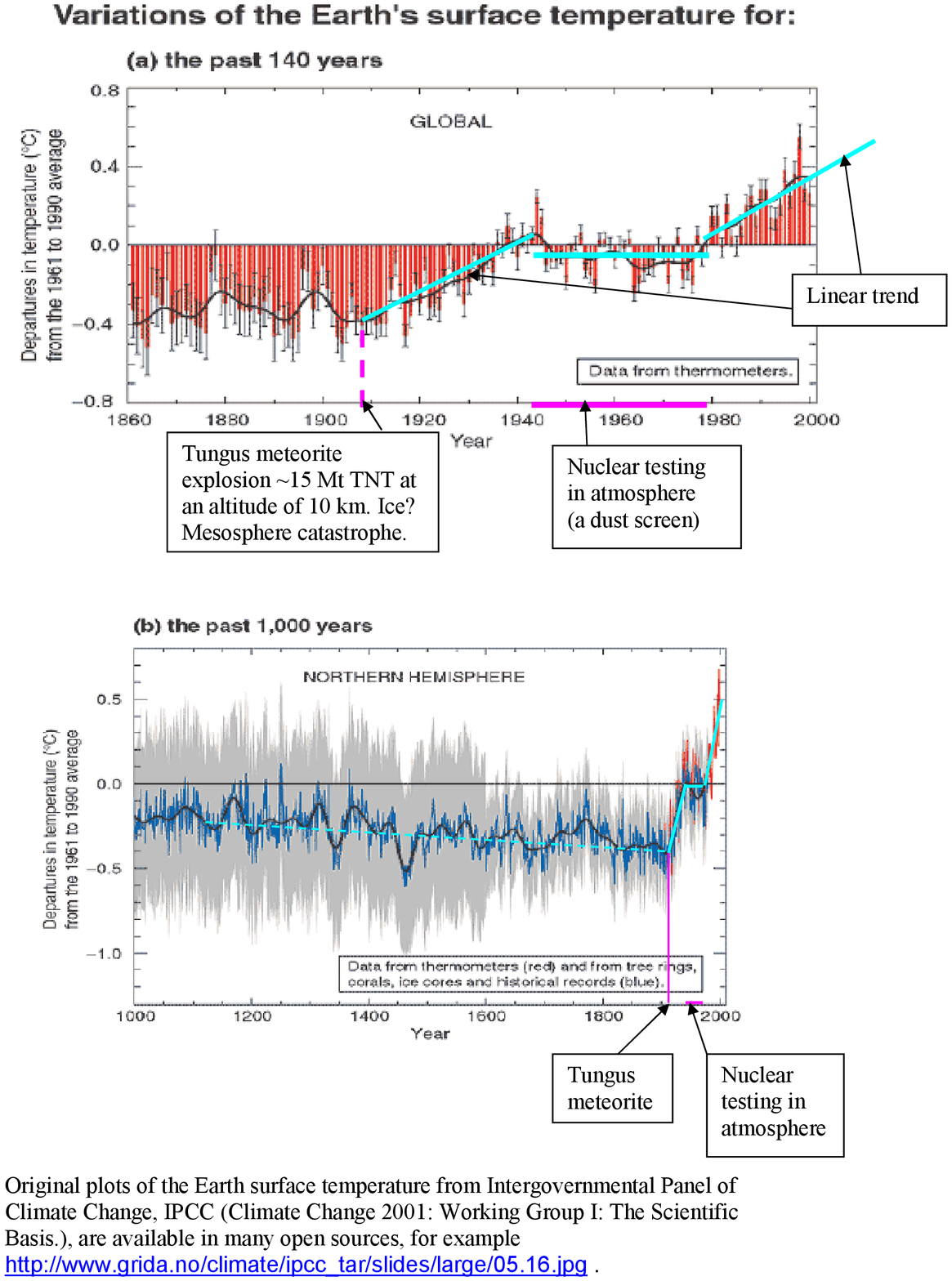}
\caption{Piecewise-linear trends.}
\end{centering}
\end{figure}

\end{document}